\documentclass[twocolumn]{aastex7}

\let\tablenum\relax
\usepackage{siunitx}
\DeclareSIUnit\arcsec{"}
\DeclareSIUnit\milliarcsec{mas}

\begin{document}

\title{Ground-Based Mid-IR Direct Imaging: The Origin of the Thermal Background on the Keck II Telescope and Correcting Instrumental Systematics}

\author[0000-0002-9242-9052]{Jayke S. Nguyen}
\email{jsn001@ucsd.edu}
\affiliation{University of California, San Diego, Astronomy \& Astrophysics Dept. 9500 Gilman Dr., La Jolla, CA 92093}

\author[0000-0002-9936-6285]{Quinn M. Konopacky}
\email{qkonopacky@ucsd.edu}
\affiliation{University of California, San Diego, Astronomy \& Astrophysics Dept. 9500 Gilman Dr., La Jolla, CA 92093}

\author[0000-0001-5684-4593]{William Thompson}
\email{William.Thompson@nrc-cnrc.gc.ca}
\affiliation{Herzberg Institute of Astrophysics, Victoria, British Columbia, CA}

\author{Natasha Popenoe}
\email{npopenoe@ucsd.edu}
\affiliation{University of California, San Diego, Astronomy \& Astrophysics Dept. 9500 Gilman Dr., La Jolla, CA 92093}

\author[0000-0003-1212-7538]{Bruce Macintosh}
\email{bamacint@ucsc.edu}
\affiliation{University of California Observatories, 1156 High Street, Santa Cruz, CA 95064, USA}

\correspondingauthor{Jayke S. Nguyen}
\email{jsn001@ucsd.edu}

\begin{abstract}
Mid-IR wavelengths are of particular interest to exoplanet science due to the fact they can extend the searchable parameter space to planets that are older and/or colder. However, a significant source of uncertainty at mid-IR wavelengths on ground-based telescopes is the thermal background. This background comes from blackbody radiation in the atmosphere and telescope and is therefore dependent on instrument design and atmospheric conditions. When performing imaging observations, this background manifests as a slowly varying, inhomogeneous signal throughout the image, underlying our data. Photometry at mid-IR can greatly constrain atmospheric models but existing data are usually scarce or have significant error bars due to the difficulty of subtracting the background. Using M-band direct imaging observing sequences on NIRC2, we evaluate the thermal background of the Keck II telescope and attempt to subtract the background in a more comprehensive way. For our primary science target, the forming protoplanet AB Aur b, we present a contrast upper limit of $2 \times 10^{-4}$ in M-band and address the limiting factors in our observation due to the thermal background. We determine that the origin of the systematic components of the thermal background comes from the K-mirror and find that the thermal background is also strongly influenced by emission from the secondary spiders on Keck II.
\end{abstract}


\section{Introduction} \label{sec:intro}
High-contrast direct imaging is one of the best methods for detailed characterization of exoplanets. By measuring the photons directly, direct imaging can probe the exoplanet's atmosphere, telling us about composition of the planet and, therefore, the formation pathway. Most of the current population of directly imaged exoplanets are biased towards massive planets (with a few exceptions) that are young and still glowing with the heat of formation. These planets are most accessible at NIR wavelengths; however, as these planets age, they cool down and their peak emission shifts to mid-IR wavelengths. Cold start planets on the other hand start with comparatively lower temperatures, peaking strongly in the mid-IR since they formed. Thus by extending to mid-IR wavelength ranges we can probe into a population of exoplanets that are ``older and colder''.

The primary issue with imaging at mid-IR wavelengths from the ground is the presence of the ``thermal background''. While most NIR wavelength observations have some background present, the background in NIR can be largely removed during the calibration phase of the observation (specifically, flats or sky frames). In the mid-IR, we begin to detect the blackbody emission of the environment, manifesting as the thermal background which cannot be removed as easily.

The thermal background is a quasi-static, spatially inhomogenous, and time-dependent structure present in ground-based imaging data. Since it changes over time, the usual calibration schemes cannot effectively remove it. The thermal background is thought to originate from three blackbody sources: the atmosphere, telescope, and instrument. The atmosphere contributes via blackbody emission from the air that we are seeing through, which can be from inside or outside the dome. The telescope and instrument can contribute through the thermal emission from the optics, telescope structure, or dust and debris. Stray light paths also likely play a larger role, since thermal emission comes from all objects that are not cooled.

Because of the difficulty associated with ground based mid-IR imaging, most exoplanet direct imaging surveys are conducted in JHK- bands (1.2 - 2.1$\mu$m) \citep{nielsenGeminiPlanetImager2019, chomezSPHEREInfraredSurvey2025, tobinDirectimagingDiscoverySubstellar2024a, liKeckHGCAPilotSurvey2024}, with few at wavelengths at or longer than L'-band (3.8 $\mu$m) \citep{skemerLEECHExoplanetImaging2016a, barcucciLIStENBandImaging2021}. In fact, most purpose-built direct imaging instruments are limited to wavelengths less than 2.5 $\mu$m; however, more general purpose imaging instruments such as NIRC2 on the Keck II 10-meter telescope, have the capability to extend out to longer, mid-IR wavelengths ($> \SI{4.5}{\micro\meter}$).

Although much of the thermal background noise can be avoided with a cryo-cooled, space-based telescope such as JWST, space-based telescopes are typically smaller than their ground-based counterparts. High angular resolution is necessary for direct imaging searches since the current demographics of Jupiter-mass planets imply that there are more planets closer-in to their host star, rather than further out \cite{nielsenGeminiPlanetImager2019}.

Despite the thermal background structure issue being well-known in mid-IR ground-based imaging \citep{kaeuflSkyNoiseMeasurement1991, otarolaAtmosphericTransmissionThermal2015, galicherBANDIMAGINGHR2011}, there is an absence of robust post-processing techniques for mitigating the thermal background. Techniques such as background LOCI \citep{galicherBANDIMAGINGHR2011} and background PCA \citep{hunzikerPCAbasedApproachSubtracting2018} work as post-processing solutions, but there is little information about the precise origin of the thermal background for each imaging system such that we can develop a more advanced methods for mitigating it.

In this paper, we present a new analysis of NIRC2 $M$-band imaging data of the AB Aur protoplanetary disk showing that the K-mirror coupled with emission from the secondary spiders comprises much of the time-dependent thermal background signal on the Keck II telescope. We additionally evaluate post-processing solutions to try and mitigate the K-mirror signal and present suggestions for observing strategies at mid-IR wavelengths from the ground.

\section{Target and Observations} \label{sec:obs}

AB Aurigae is a nearby Herbig Ae star, with spectral type A0. At a distance of 156 pc, it hosts a very well-studied protoplanetary disk that has an accreting protoplanet candidate \citep{currieImagesEmbeddedJovian2022}. The proximity of the system and size of the disk mean that the system is relatively well spatially resolved, and has an angular size of approximately $2" \times 2"$. It has been probed in NIR wavelengths and with polarimetric imaging but lacks M-band data because of the challenges associated with the mid-IR thermal background. The faint protoplanet candidate has yet to be seen at longer wavelengths \citep{marinasMidInfraredImagingHerbig2006, jorqueraLargeBinocularTelescope2022a, speedieMappingMergingZone2025b}.

\begin{deluxetable}{c|c|c|c|c|c|c|c}
\tablehead{
\colhead{UTC Date} & \colhead{Integration Time (s)} & \colhead{Seeing (")} & \colhead{Avg. Strehl (\%)} & \colhead{\# Sci Frames} & \colhead{\# Sky Frames} & \colhead{Dithering} & \colhead{K-mirror}
}
\startdata
2024-02-20 & 4180.8 & 0.68 & 61.28 $\pm$ 2.97 & 140 & 24 & Y & Y \\
2024-10-24 & 4586.6 & 0.48 & 57.39 $\pm$ 5.99 &  149 & 10 & N & N \\
2024-10-25 & 3432.4 & 0.54 & 63.08 $\pm$ 13.71 & 114 & 33 & N & Y \\
2024-10-26 & 2243.4 & 0.62 & 61.49 $\pm$ 6.48 & 124 & 43 & N & Y \\
\enddata
\caption{Summary of our observations. Nightly seeing measurements are averages obtained from the Mauna Kea Weather Center Differential Image Motion Monitor (DIMM). The DIMM is a wavefront sensor that measures the Fried parameter and the atmospheric seeing and is located near the Keck II telescope \citep{cherubiniModelingOpticalTurbulence2008a}.}
\label{tab:observations}
\end{deluxetable}

\begin{deluxetable}{c|c|c|c|c|c|c}
\tablehead{
\colhead{} \vspace{-0.25cm} & \colhead{Outside} & \colhead{Inside} & \colhead{Inside} & \colhead{Outside} & \colhead{Primary} & \colhead{Secondary} \\
\colhead{UTC Date} & \colhead{Temperature (C)} & \colhead{Temperature (C)} & \colhead{Humidity (\%)} & \colhead{Humidity (\%)} & \colhead{Temperature (C)} & \colhead{Temperature (C)}
}
\startdata
2024-02-20 & $3.00  \pm 0.18 $ & $2.18  \pm 0.14 $ & $0.69  \pm 0.17 $ & $0.40  \pm 0.14 $ & $0.82  \pm 0.17 $ & $0.82  \pm 0.20 $ \\
2024-10-24 & $4.50  \pm 0.24 $ & $4.10  \pm 0.14 $ & $9.18  \pm 1.67 $ & $8.50  \pm 1.35 $ & $4.06  \pm 0.09 $ & $2.43  \pm 0.11 $ \\
2024-10-25 & $4.00  \pm 0.14 $ & $3.50  \pm 0.14 $ & $34.21 \pm 1.25 $ & $31.63 \pm 1.39 $ & $3.73  \pm 0.05 $ & $1.90  \pm 0.20 $ \\
2024-10-26 & $3.20  \pm 0.14 $ & $3.20  \pm 0.07 $ & $32.66 \pm 2.46 $ & $31.54 \pm 2.95 $ & $3.39  \pm 0.03 $ & $2.33  \pm 0.17 $ \\
\enddata

\caption{Median temperature and humidity values and their standard deviation for each of the observations. Data are collected with sensors co-located with the Keck II dome. Primary and secondary temperatures are obtained from sensors placed on the primary and secondary mirror of the telescope. This ancillary weather data was obtained from the from the Keck Observatory Archive.}
\label{tab:weather}
\end{deluxetable}

Our data is comprised of four M-band ($\lambda = \SI{4.67}{\micro\meter}$, $\Delta\lambda=\SI{0.24}{\micro\meter}$) half-nights of data on the AB Aur system, using NIRC2 on Keck II on UT 2024-02-20, 2024-10-24, 2024-10-25, and 2024-10-26. The primary goal of the observations was to observe the disk at long wavelengths and obtain an M-band photometric datapoint on the candidate protoplanet. Previous studies on this protoplanet have data extending to L'-band \citep{currieImagesEmbeddedJovian2022}, but new models suggest that the enshrouded protoplanet is expected to have a peak flux at M-band \citep{choksiSpectralEnergyDistributions2025a}.

We use NIRC2\footnote{https://www2.keck.hawaii.edu/inst/nirc2/} \texttt{narrowcam} mode which has a 10"$\times$10" field of view across 1024$\times$1024 pixels, corresponding to a pixel scale of $\sim$0.01" per pixel. For consistency, all frames had an individual integration time of \SI{0.104}{\second} with 300 coadds, subarrayed to a region of 768$\times$776 pixels, with one exception that during the 2024-10-26 epoch, we used a 512$\times$512 pixels subarray of the detector. Each of our observations used the \texttt{largehex} pupil mask, which is an oversized hexagonal obscuration of the secondary, and a very slightly oversized mask of the spiders (Figure 5, \cite{cosensLigerNextgenerationKeck2020}). We used vertical angle mode for angular differential imaging, except for one epoch, 2024-10-24 where we used stationary angle mode. A tabular summary of our observations can be found in Table \ref{tab:observations} and a summary of the weather conditions can be found in Table \ref{tab:weather}.

The first dataset on 2024-02-20 was taken using an observing strategy optimized for background LOCI. We remained on target in roughly 30 minute intervals, and at the end of each interval, we dithered off target to blank sky to take sky frames for approximately 15 minutes. We then returned to the target, placing it in a different region of the detector. The atmospheric seeing throughout the night was clear and humidity was exceptionally low during this particular epoch.

On 2024-10-24, we tested an experimental observing strategy in which we turned off the instrument de-rotator. Additionally, to test the effects of the AO system on the background, we collected open-loop and closed-loop data.

In vertical angle mode used commonly in classical ADI \citep{maroisAngularDifferentialImaging2006a} the K-mirror is not completely stationary. The K-mirror needs to rotate to account for the rotation of the telescope pupil due to changes in target elevation. The additional rotation due to elevation comes from the geometry of the optical system since our instrument is mounted on the Nasmyth deck of an alt-az telescope. In our attempt to mitigate the effects of the K-mirror, we turn off the rotation completely and allow the PSF to rotate in the frame.

Angular differential imaging is still possible in this case since the PSF signal and planet signal are still rotationally separated, with the final post-processing de-rotation step requiring an additional de-rotation by the elevation angle to align the PSF, after which we can apply typical PSF subtraction techniques. 

Observations on 2024-10-25 and 2024-10-26 used a simpler observing strategy than the previous epochs, placing the object on one location on the detector, with sky frames taken in roughly 30 minute intervals. Frames that were poor quality due to poor weather conditions and high wind speeds were removed from the final observing sequence. Both nights, the temperature and humidity were relatively stable throughout the evening. However, the sky was relatively cloudy during both epochs and many frames had to be rejected due to cloud coverage.

We follow standard astronomical data reduction techniques of subtracting darks and flat fielding. Flat fielding however, takes a few extra considerations in M-band. Typically, flats are taken using a dome lamp inside of the telescope, however the thermal emission from within the dome is too bright and quickly saturates the image even at the shortest exposure times. Dome flats are therefore not possible for flat fielding in M-band. Additionally, using flats at other wavelength bands does not work since the flat field response of the detector is significantly different in M-band. Instead, we use sky flats during our observing sequences. Sky flats use the relatively uniform emission of the sky to flat field the frames. We also match the coadds and integration time of the science frames for consistency. To obtain a sky frame, we dither a few arcseconds away such that there are no sources the frame. The sky frame is approximately the same altitude and azimuth of the target. To make the master flat frame we take all of the sky frames throughout the night, subtract them by their respective median to account for differences in elevation, and stack them. We then median combine the stack, add an offset to remove negative values, then divide by the median value of the combined image. Since we take sky flats periodically throughout our observations, the thermal background is smeared and averaged as the K-mirror rotates, smoothing out the thermal background between frames. Thus, with this sky flat fielding technique, we only capture the effects of pixel-to-pixel variation in the detector instead of any other sources like the thermal background.

\section{Thermal Background Analysis} \label{sec:derotator}

\begin{figure*}
    \centering
    \includegraphics[width=\linewidth]{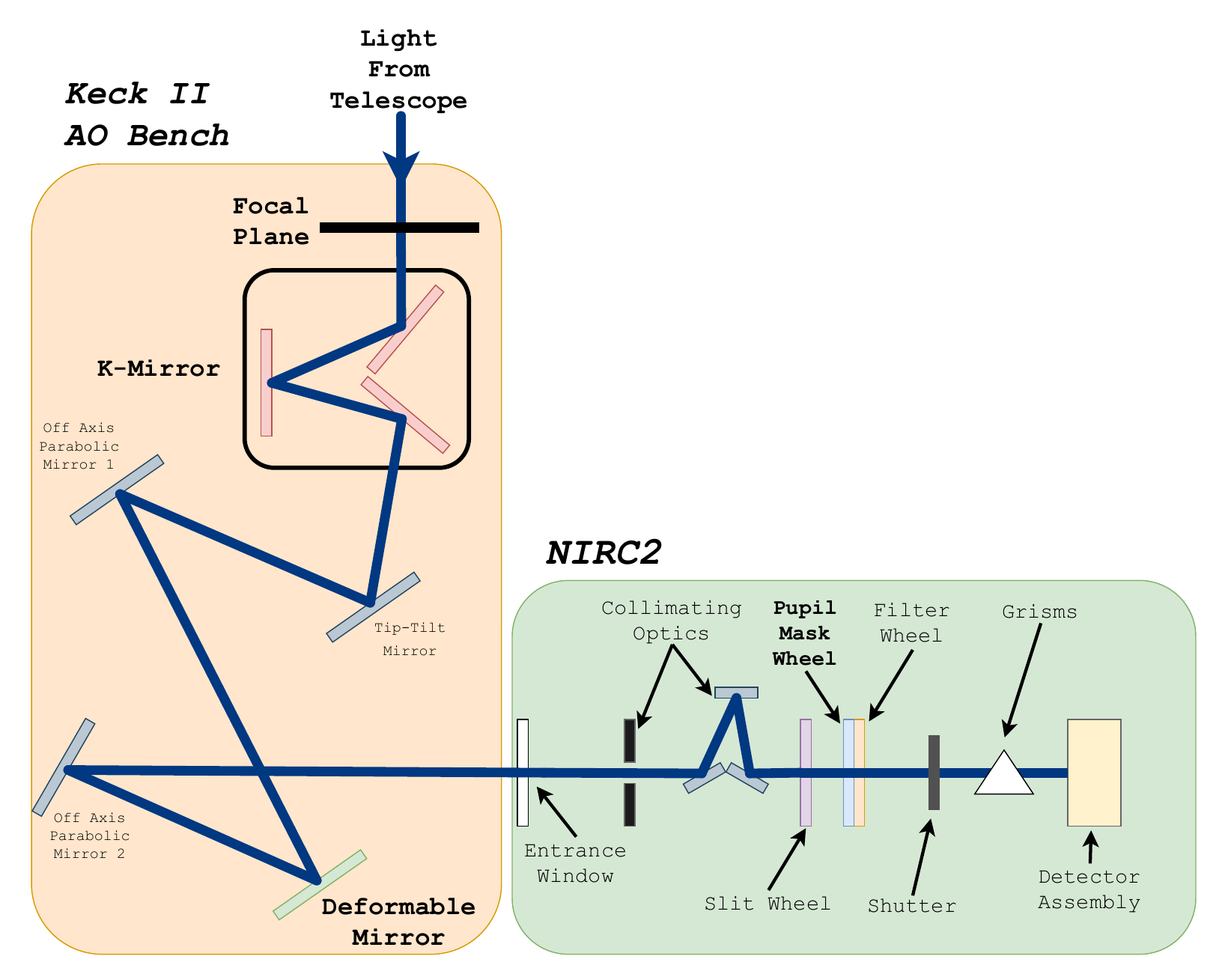}
    \caption{The optical diagram of the Keck II adaptive optics bench and the NIRC2 instrument. The Keck II AO bench is not cooled and is slightly warmer than ambient due to the electronics, while the entirety of NIRC2 is contained inside a cooled dewar. The first focal plane is just before the K-mirror. Optics that are most relevant to this study are bolded. This figure is adapted from figures from the NIRC2 Observer's Manual (https://www2.keck.hawaii.edu/inst/nirc2/) and Figure 1 from \cite{lilleyNearinfraredPyramidWavefront2018}. The optical layout is roughly to scale, but is only illustrative.} 
    \label{fig:optical}
\end{figure*}

\begin{figure}
    \centering
    \includegraphics[width=\linewidth]{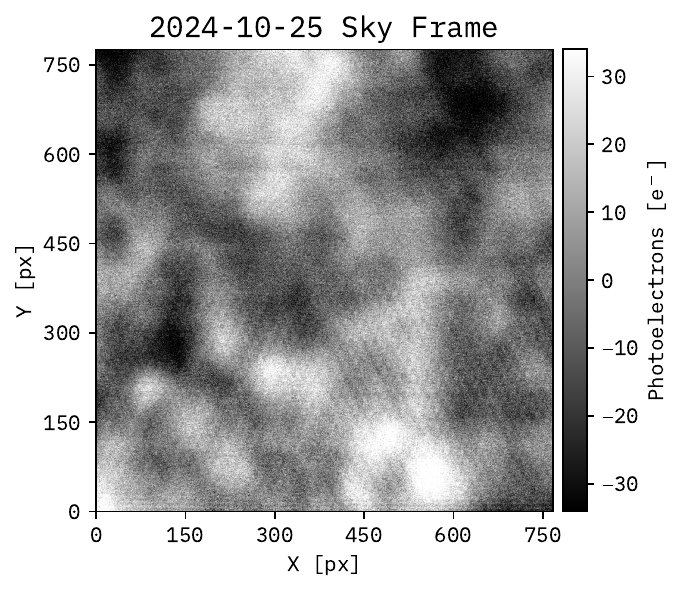}
    \caption{A median subtracted sky frame from our 2024-10-25 observing night that shows the typical spatial extent and intensity of the thermal background.}
    \label{fig:sky_frame}
\end{figure}

\begin{figure}
    \centering
    \includegraphics[width=\linewidth]{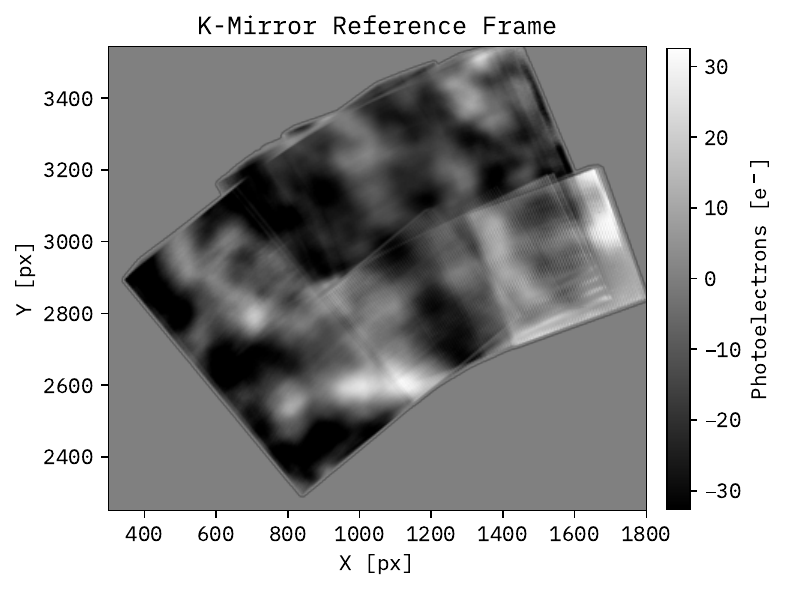}
    \caption{The K-mirror map in the K-mirror reference frame using science data from 2024-02-20. The frames were processed with an affine transform that performs a translation and rotation, transforming them into the K-mirror reference frame. The PSF in the science frames is masked out and we average over the overlapping regions of the data cube.}
    \label{fig:derotator_fig}
\end{figure}

The main contributions to the thermal background come from three primary components: emission from within optical system of the telescope (Figure \ref{fig:optical}), the emission from telescope structure, and the sky brightness of the atmosphere. The relative strength of each of these components varies depending on the particular conditions of the observation.

An example of the thermal background component present in our imaging data is shown in Figure \ref{fig:sky_frame}. In general the thermal background has a structure, with extended features that persist between frames. These features are not random and are fairly inhomogeneous across the frame. The background evolves slightly between frames, and no two frames are exactly alike.

\subsection{The K-mirror}

Previous studies such as \cite{galicherBANDIMAGINGHR2011} have alluded to the K-mirror as a significant source creating the features seen in the thermal background; however, exactly how it influences the thermal background on images is unknown. Figure \ref{fig:optical} shows the location of the K-mirror in the optical path of the Keck II AO bench. The main function of the K-mirror is to rotate an image by a particular angle to keep north-up in images, or to track the telescope pupil (as with angular differential imaging) on an alt-az telescope with instruments at the Nasmyth focus. Critically, the K-mirror has the property that the desired image rotation angle is half of the drive angle of the K-mirror. That is, if you want to rotate your image by angle $\theta$, you only need to rotate your K-mirror by angle $\theta/2$. This means that if the K-mirror is the source of the thermal emission, it would rotate at a different angular speed than the parallactic angle, meaning that the K-mirror signal is rotationally separate from both the PSF and rotating planet signal. Additionally, the K-mirror is the first optic on the AO bench after the telescope optics and is located near the first focal plane of the system.

\subsection{K-mirror Reference Frame}

In reduced science images, there appears to be a rotating component in the residual background of the image. Assuming that this component comes from the K-mirror, if we de-rotated by the K-mirror angle, the thermal background would become static in this rotating reference frame. However, the NIRC2 \texttt{narrowcam} mode is offset by \SI{13.75}{\arcsec} from the optical axis, so transforming into this reference frame requires both a translation and rotation. 

Testing this, we perform a translation by this value and rotation by half of the K-mirror drive angle; and indeed this tracks the thermal background as it appears in reduced data, as shown in Figure \ref{fig:derotator_fig}. The offset is an estimation based off of instrument specifications and the actual location of the optical axis is likely slightly different. This estimation is accurate enough for demonstrating this effect, however.

We speculate that part of the thermal emission is dust on the surface of the K-mirror or optical imperfections. Its location near a focal plane in the system means that dust sitting on the surface of the K-mirror is most readily ``imaged'' since the dust on the surface are sources emitting as a blackbody. The dust features imaged are unlikely to be single individual dust particles, but more likely a mottling on the mirror surfaces. Furthermore, variation between frames and between epochs preclude any kind of consistent correction in the form of a static ``dust map''.

\subsection{Open Loop Thermal Background Observations}

\begin{figure}
    \centering
    \includegraphics[width=0.975\linewidth]{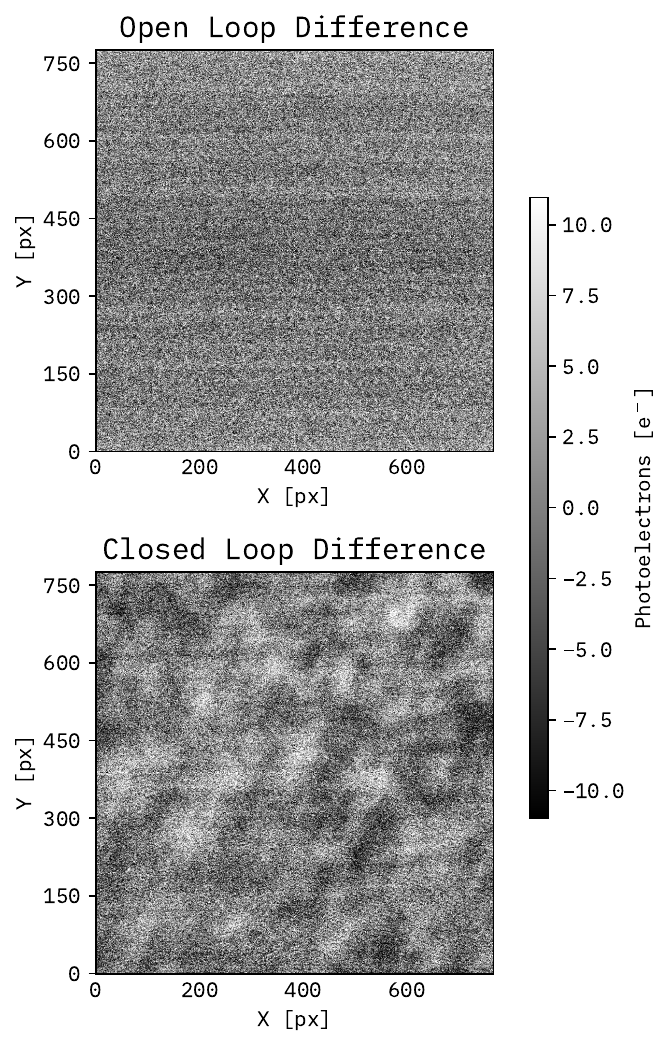}
    \caption{The difference between consecutive sky frames from the 2024-10-24 epoch in open loop and closed loop modes of the adaptive optics system.}
    \label{fig:open_loop_data}
\end{figure}

During the 2024-10-24 epoch, we took sky frames while turning the adaptive optics system on and off. We find that when the loop is closed, the thermal background appears to fluctuate between frames, whereas with the loop open, the thermal background is largely static. This effect can be seen clearly in Figure \ref{fig:open_loop_data}. By taking the difference between consecutive frames, we can measure the relative change between frames. In the open-loop case, we have a very uniform difference and in the closed-loop case we see large spatial residuals indicating significant changes between frames.  The movement we are seeing is due to the deformable mirror in the adaptive optics system modulating the signal from the thermal background. This modulation makes it so that the thermal background is always slightly different between two frames in a quasi-random way.

To understand the modulation, consider a high-emissivity point source (such as a small coating defect or large dust particle) at or near the telescope focal plane. In a simple camera, thermal emission from that location would produce a bright point in the final image. In an adaptive optics instrument, the light path includes the deformable mirror. When the adaptive optics loop is closed, the deformable mirror is assuming a time-varying shape that is the conjugate to the atmospheric turbulence above the telescope. The wavefront therefore experiences a distortion that is roughly equal (but opposite) the the distortion a wavefront experiences as it propagates through the Earth's atmosphere. As a result, the point is blurred into an extended image similar to the atmospheric PSF. An extended pattern of emission near a focal plane (such as the emission from the K-mirror) will be convolved with this internal PSF, smoothing it out. This results in a different thermal emission background with AO active or inactive. Even with AO active, as the seeing fluctuates, and as AO performance varies, the internal PSF will vary.

\subsection{The Photon Noise Limit}

In direct imaging observations, the deepest contrast that we can achieve is limited by two factors: the brightness of the target star and the noise level of our detector. Typically, most of the random noise in the signal comes from shot noise, the read noise, and the dark current. NIRC2 has a very low dark current (\SI{0.1}{e^-/px/sec}), so we neglect this term. Additionally, at thermal wavelengths we have an additional background signal, which adds both random shot noise and a systematic noise term. Examining the noise only due to the thermal background and detector limitations, the equation for noise in our system is Equation \ref{eq:noise}:

\begin{equation}
\sigma_\text{total}^2 =  \underbrace{\sigma_\text{shot,thermal}^2  + \sigma_\text{read}^2}_\text{Random} + \underbrace{\sigma_\text{thermal}^2}_\text{Systematic}
\label{eq:noise}
\end{equation}

The photon noise limit of our observations is set only by the random component of this equation. The systematic term, in theory, can be minimized by careful processing of the data.

The shot noise due to the thermal background varies depending on the airmass of the observations. For each of the nights, we have a median value of 8000-15000 $e^-$, and 300 coadds per frame. As a conservative estimate, we assume the greater value. Using Poissonian statistics, we obtain: $\sigma_\text{shot,thermal} = \sqrt{15000/300} = 7.07$ $e^-$.  NIRC2 has a read noise of $50$ $e^-$ per frame, but averaged over 300 coadds this becomes $\sigma_\text{read} = 50/\sqrt{300} = 2.88$ $e^-$.

Adding these two terms in quadrature, we can obtain the random component of our noise, the photon noise limit:

\begin{equation}
\sqrt{\sigma_\text{read}^2 + \sigma_\text{shot,thermal}^2} = 7.63 \text{ }e^-
\label{eq:random}
\end{equation}

This value is our lowest possible error estimate needed to understand our sensitivity. To reach the photon noise limited regime, we need to minimize the contribution of $\sigma_\text{thermal}$.

\subsection{Thermal Background Properties}

\begin{figure}
    \centering
    \includegraphics[width=\linewidth]{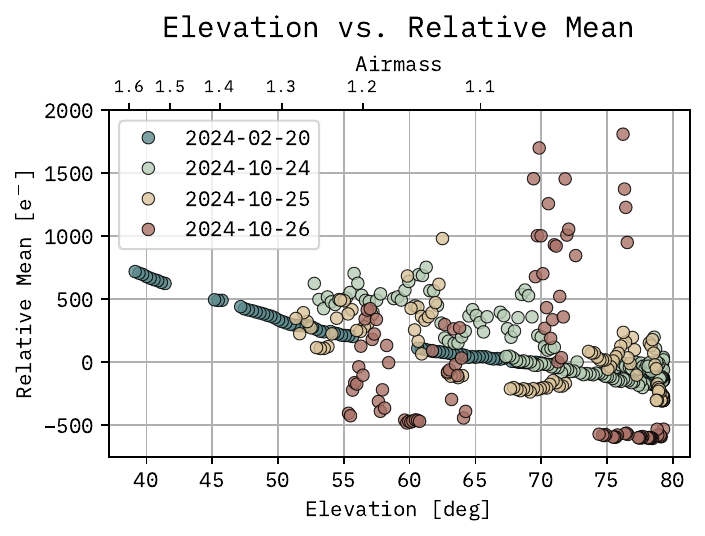}
    \caption{The ``relative mean'' versus elevation. The relative mean is the average value of each frame, subtracted by the minimum value of each epoch. This gives the relative change in the average value as a function of elevation.}
    \label{fig:elevation_v_relativemean}
\end{figure}

\begin{figure}
    \centering
    \includegraphics[width=\linewidth]{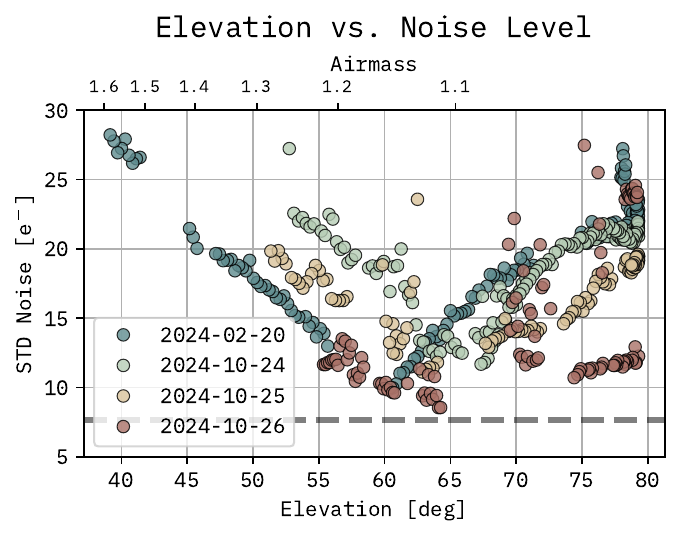}
    \caption{The standard deviation in the image versus the elevation. The gray dashed line is the theoretical photon noise limit as found in Equation \ref{eq:random}. The noise was calculated by taking the mean-subtracted pixel values of the background in each of the science frames masking out the PSF and plotted against the pointing elevation.}
    \label{fig:elevation_v_noise}
\end{figure}

\begin{figure*}
    \centering
    \includegraphics[width=\linewidth]{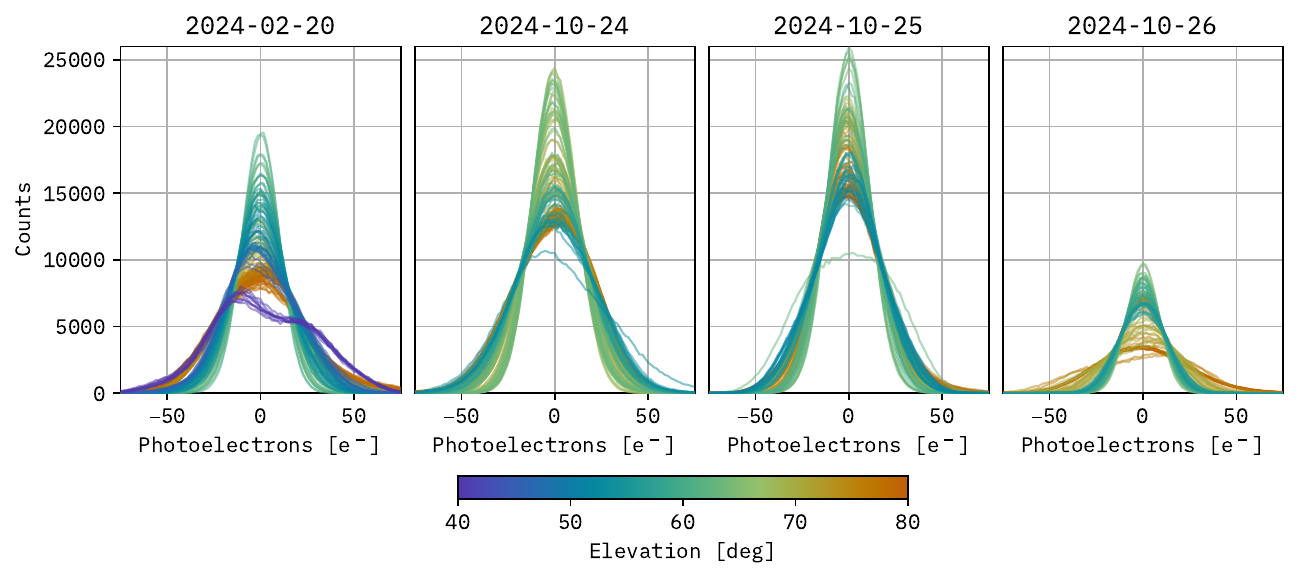}
    \caption{The histogram of the mean-subtracted pixel values of each of the science frames, where each line is a different frame for the given epoch. Each histogram has 100 bins ranging from -75 e- to 75 e-. The PSF was masked out with a circle 200 px in radius prior to binning. The pointing elevation of the telescope is denoted by the color of plot.}
    \label{fig:obs_hist}
\end{figure*}

The thermal background adds a significant offset in the mean value of the background in our image. Since the thermal background signal is relatively strong, it limits our overall dynamic range since we need to use short integration times so that we do not reach the non-linear regime of the detector. These shorter integration times mean that we incur an additional read noise penalty since we have to read out comparatively more frames, in addition to reducing the on-target efficiency of our observation due to the readout time overhead. Coadding large groups of frames helps average away random noise components, but a side-effect is that we average the PSF over time, and have fewer PSF realizations to give to our PSF subtraction algorithms.

Measuring the offset of the thermal background, we find that the mean value of the background decreases as elevation increases, as shown in Figure \ref{fig:elevation_v_relativemean}. This is expected as the airmass decreases as we point the telescope higher in elevation. Note that cloudy observing conditions for the 2024-10-26 epoch make this relationship less clear, but the linear decrease in the mean value is apparent in all of the observation epochs.

When measuring the standard deviation of the thermal background, we find that the noise, $\sigma_\text{thermal}$, is not a simple function of elevation. In Figure \ref{fig:elevation_v_noise} we find that the noise decreases with increasing elevation until reaching a particular elevation angle, then increases as elevation rises. For the purposes of this paper, we will call this the noise inversion. The elevation where this occurs varies slightly between nights. We also find that the measured noise for most elevations is higher than the photon noise limit as found in Equation \ref{eq:random}, leaving a significant margin for improvement.

\subsection{The Noise Inversion}
The noise inversion is observed during all four observing nights, despite different observing conditions. The observed increase in noise as a function of elevation is interesting since intuitively, one would expect that decreasing airmass would lead to lower thermal background overall if the atmosphere were the main contributor. If the K-mirror were the only contributing factor to the background we would expect that the noise levels in the background stay consistent over time, only changing as a function of air temperature or humidity. Additionally, the K-mirror was set to a fixed, user-specified angle on 2024-10-24, and this date still shows increasing noise at high elevations. This leads to the conclusion that the neither atmosphere nor the K-mirror are driving this behavior; rather, something systematic within the telescope or instrument(s) is the source of the noise inversion.

The noise increase begins at an elevation that is still relatively shallow, and its unlikely that stray light or reflections from within the dome can enter into the optical path given the baffles on both the secondary and tertiary mirrors on Keck. On the telescope structure however, the spiders supporting the secondary are emitters in the telescope path. Normally, these spiders are blocked by a rotating pupil mask that is cooled within the NIRC2 dewar. However, there is an optical alignment offset between the K-mirror on the Keck II AO bench that introduces a nutation into the system. As found in \cite{cosensLigerNextgenerationKeck2020}, the nutation is such that at higher elevations, the pupil mask and the secondary spiders become more misaligned. Based on Figure \ref{fig:optical}, we can see that any kind of small misalignment in the K-mirror will propagate through the optical system and will affect downstream components. We therefore hypothesize the following scenario that explains Figure \ref{fig:elevation_v_noise}: the pupil and secondary spiders are slightly misaligned at low elevation. As the elevation increases the pupil mask slowly starts to cover more of the secondary spiders as alignment improves until reaching a maximum alignment (minimum noise) blocking out most of the emission from the spiders. Finally as elevation continues to increase the pupil mask and secondary spiders again become misaligned, reaching a maximum misalignment. We also note that the noise at around \SI{78}{\degree}, continues to increase slightly, likely due to atmospheric changes within the dome that change the emissivity of the spiders, or due to an azimuthal dependency of the nutation.

\begin{figure*}
    \centering
    \includegraphics[width=0.8\linewidth]{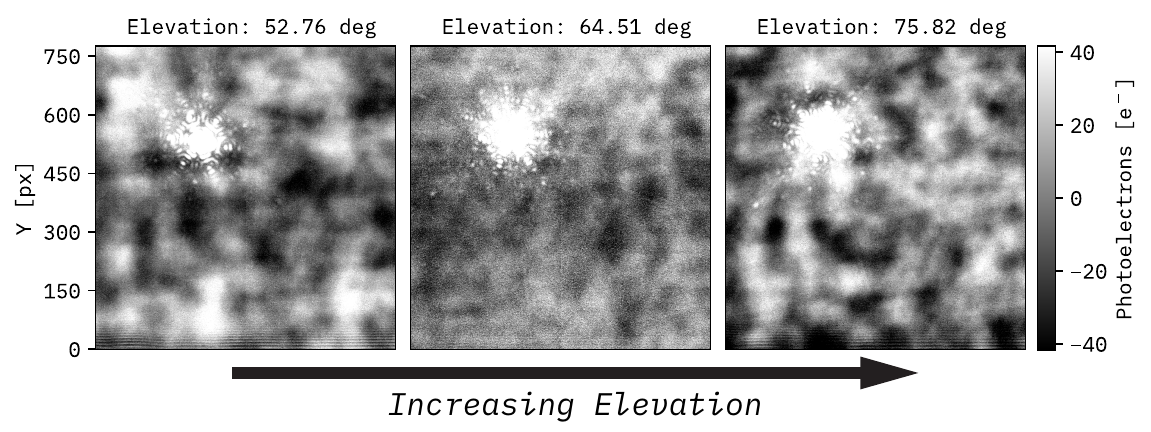}
    \caption{Median subtracted sky frames from the stationary K-mirror epoch, 2024-10-24. Shows the structure of the thermal background before, in the middle of, and after the noise inversion.}
    \label{fig:states}
\end{figure*}

Looking at the histograms as a function of elevation in Figure \ref{fig:obs_hist}, it is apparent that the actual noise content is different within the images as the telescope changes elevation. At low and high elevations, we see two different histograms, with 2024-02-20 showing a double-peaked histogram at low elevations and a single-peaked skewed histogram at high elevations. When examining the images, the structure of the thermal background changes significantly after the noise ``inverts'' at high elevations. That is, the structure in the background before the inversion (occurring between \SI{60}{\degree} to \SI{70}{\degree}) is different from the structure afterwards. This can be seen clearly in Figure \ref{fig:states}, where the thermal background looks significantly different after the inversion. We speculate that this is due to emission from the secondary support spiders acting like a source, illuminating the dust on the K-mirror at slightly different angles, resulting in different patterns on the thermal background. The noise levels during the inversion are low and still have some structure apparent in the background since we expect that there is some misalignment even at the minimum noise levels.

\subsection{Weather Impact}

Looking at the weather data from Table \ref{tab:weather}, there are no significant large-scale changes in the weather patterns that are correlated with the spatial structure of the thermal background. Cloud cover changes the overall median levels of the background, as seen in Figure \ref{fig:elevation_v_relativemean}, but the actual structure of the thermal background was relatively unchanged as clouds passed overhead. This is likely because we are integrating over a relatively large air column which serves to smooth the overall background emission. Moreover, the sky background is an extended emission in the seeing air column starting at pupil plane, meaning that it can only contribute a smooth overall background in the focal plane. Additionally, from a geometric perspective, stray light from the atmosphere beyond the atmospheric seeing column is also unlikely to reach our instrument. Building on our model of the thermal background, we find that the atmospheric component is a relatively simple component of the thermal background, only changing the median levels as a function of airmass.

\begin{figure*}
    \centering
    \includegraphics[width=0.75\linewidth]{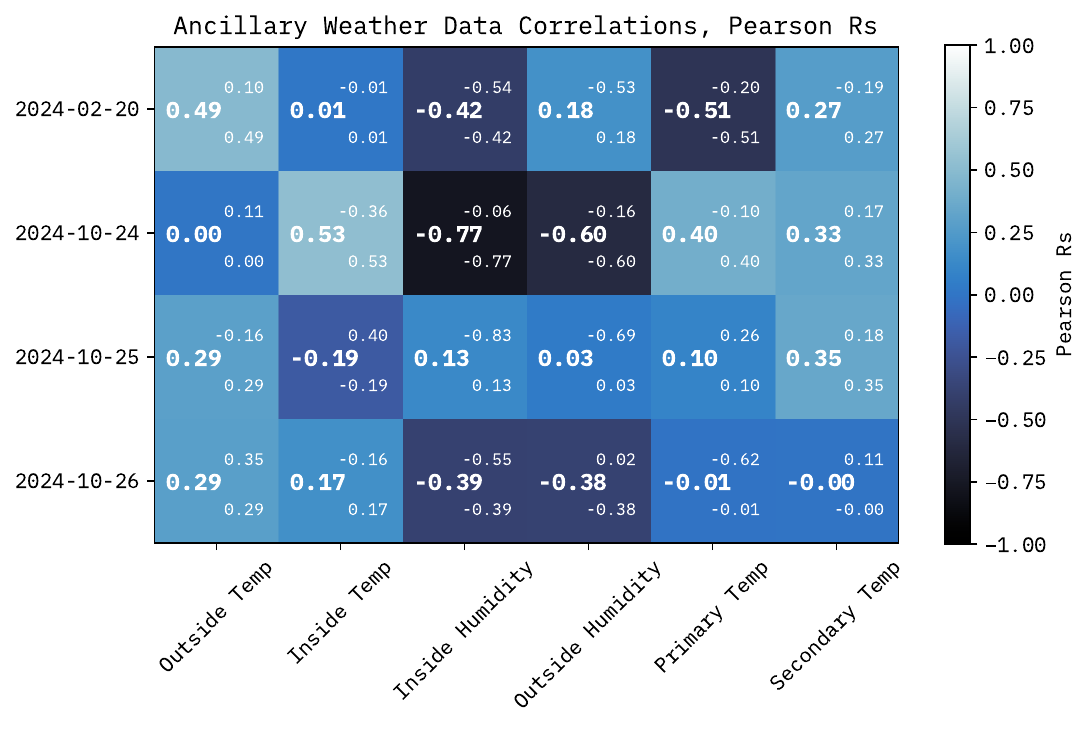}
    \caption{A correlation matrix of ancillary weather data from the Keck Observatory Archive for each of our observations against the noise levels of the epoch. We calculate the Pearson correlation coefficient for all of the data. The r-values are quoted with the upper and lower confidence intervals, where the color corresponds to the r-value itself. The ``outside'' values are obtained from sensors nearby outside the Keck II dome, while ``inside'' values are obtained from locations within the dome. The primary temperature is obtained with a temperature sensor near the primary mirror, and the secondary temperature is obtained with sensors near the secondary mirror.}
    \label{fig:ancillary_weather_data_correlations}
\end{figure*}

To evaluate the weather on a smaller scale, closer to the telescope, we obtained ancillary weather data from sensors placed inside and outside the Keck II dome from the Keck Observatory Archive. To check for correlations, we calculate the Pearson correlation coefficient between the noise levels in the data and the ancillary weather data. Looking at the results in Figure \ref{fig:ancillary_weather_data_correlations}, we could not find any consistent correlations with the noise levels in the thermal background and the various temperature and humidity sensors placed around the Keck II telescope, inside and outside the dome. In one case, there seems to be consistent correlations with the outside temperature and secondary temperature, but the temperature varied by less than a degree throughout each night as shown in Table \ref{tab:weather}. This is consistent with the findings of \cite{sauterDetectionLimitsThermalinfrared2024a}, which looked at Keck data in L'-band.

\begin{figure}
    \centering
    \includegraphics[width=\linewidth]{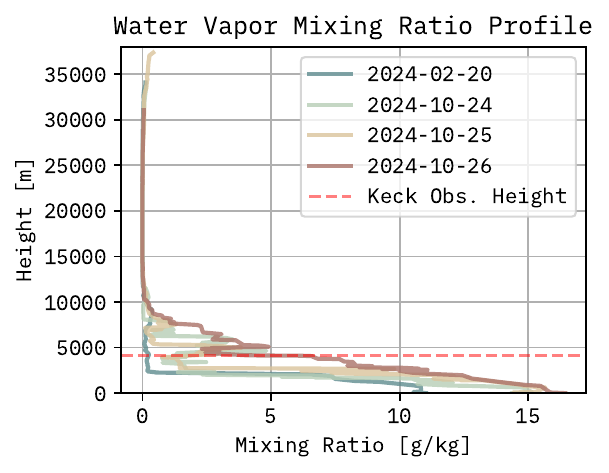}
    \caption{Atmospheric mixing ratio profile indicating the grams of water present per kilogram of air in the atmosphere. These data were collected from a weather balloon released from Hilo at 2AM HST the night of the observations.}
    \label{fig:atmo_mix_profile}
\end{figure}

\begin{figure}
    \centering
    \includegraphics[width=\linewidth]{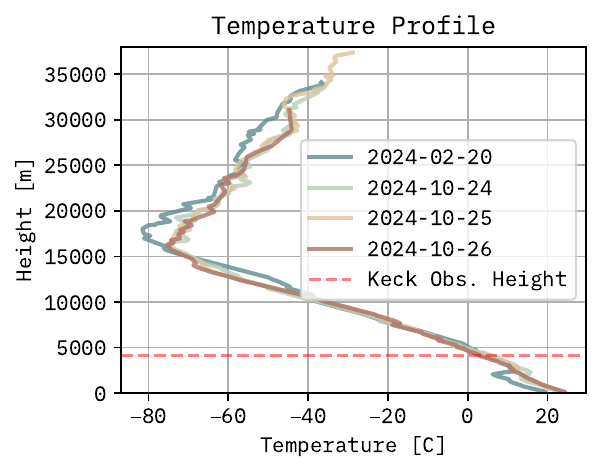}
    \caption{Atmospheric temperature profile obtained from radiosonde data of a weather balloon released from Hilo at 2AM HST the night of the observations.}
    \label{fig:atmo_temp_profile}
\end{figure}

Looking beyond the immediate weather environment, we analyze radiosonde data from the University of Wyoming Atmospheric Science Radiosonde Archive that was collected from a weather balloon that is launched daily from Hilo. This data obtained for each of the observing nights tells a similar story.

\subsubsection{Atmospheric Absorption}
Water vapor content may impact our observation since it is a significant absorber in the M-band wavelength range. Less water vapor along the seeing column of air will result in less atmospheric absorption and vice versa. While there are other absorbing molecules in M-band, water vapor is the only one present at significant levels that can affect our photometry. In Figure \ref{fig:atmo_mix_profile} we can see that each night had fairly consistent water vapor content with layers slightly above the telescope, except for 2024-02-20. In 2024-02-20, there was significantly less water vapor than the previous epochs, but the exact impact of this is difficult to quantify, since we later find that all of the epochs when reduced were of roughly similar data quality. Additionally, as previously discussed, our measurements are ultimately limited by the internal thermal background rather than absorption features.

\subsubsection{Atmospheric Emission}
From Figure \ref{fig:atmo_temp_profile} we can see that the general atmospheric temperature profile was fairly consistent between epochs. The M-band atmospheric background emission comes from the short-wavelength side of the blackbody distribution. This means that as the temperature of the air gets cooler, the emission decreases. The density of the air decreases exponentially with altitude, resulting in very little emitting power in warmer air at high altitudes. Looking at both the temperature profile and water vapor content, we see that most of the blackbody emitting power of the atmosphere is in the layers of air closest to the telescope. The relatively long focal length of the telescope ensures that these nearby layers of air will always be out of focus. The atmosphere thus cannot form images consistent with the spatial structure that comprises most our systematic error in the thermal background.

\section{Thermal Background Removal and Post-Processing}

We add an additional background subtraction step for post-processing M-band data. After flat fielding, there is still a very significant thermal background component underlying our data, as expected.

We tested three approaches for the removal of the thermal background: a simple median subtraction of the background, background LOCI as described by \cite{galicherBANDIMAGINGHR2011}, and a new form of model-dependent background subtraction which we term ``Extra Basis LOCI''.

In Extra Basis LOCI, we perform a least-squares minimization of the large-scale thermal background, similar to background LOCI. The motivation is that background LOCI does not take into account the model of the rotating background due to the K-mirror. Using our understanding of how the thermal background behaves, we seek to add an extra basis to remove the rotating signal. In typical background LOCI, the image basis for performing least-squares subtraction is drawn from the science frames, where the PSF is dithered to another location so that the thermal background can be measured without the influence of the PSF. Modifying this scheme, we add extra basis images on top of what is already included in background LOCI. We perform a linear transformation of science frames into the K-mirror reference frame, so that the rotating thermal background is ``tracked''. We additionally de-rotate sky frames taken later in the night, also adding them to our basis. Since we are fitting such large-scale features, we do not expect that there is significant self-subtraction from this step. If done correctly, none of the planet signal is contained in the basis frames during the background subtraction.

Due to our observing strategy, using background LOCI was not always possible since we did not dither on 2024-10-24, 2024-10-25, and 2024-10-26. Additionally, on 2024-10-24, we kept the K-mirror stationary to test whether we could adequately remove the background if it were not rotating.

After the thermal background is removed, the flux of each image is normalized to the median flux of the observing sequence. A template PSF is then made using the frames in the observing sequence, and we sub-pixel register each image using a fit to the template PSF. To perform the PSF subtraction, we use \texttt{ADI.jl} \citep{Lucas2020} on the registered sequence. We use principal component analysis (PCA), using 16 principal components. Fewer components are better considering that AB Aur b is an extended source. Using more principal components did not result in a cleaner subtraction, and using fewer components decreased the SNR. We calculate the per-epoch throughput corrected $3\sigma$ contrast curves, using the \texttt{contrast\_curve} function in \texttt{ADI.jl}. Furthermore, we modify \texttt{ADI.jl} to perform a multi-epoch median combination of the images, calculating a multi-epoch throughput corrected contrast curve by injection recovering the same model companion throughout all epochs, weighting each epoch by the total integration time.

\section{Results}

\begin{figure*}
    \centering
    \includegraphics[width=\linewidth]{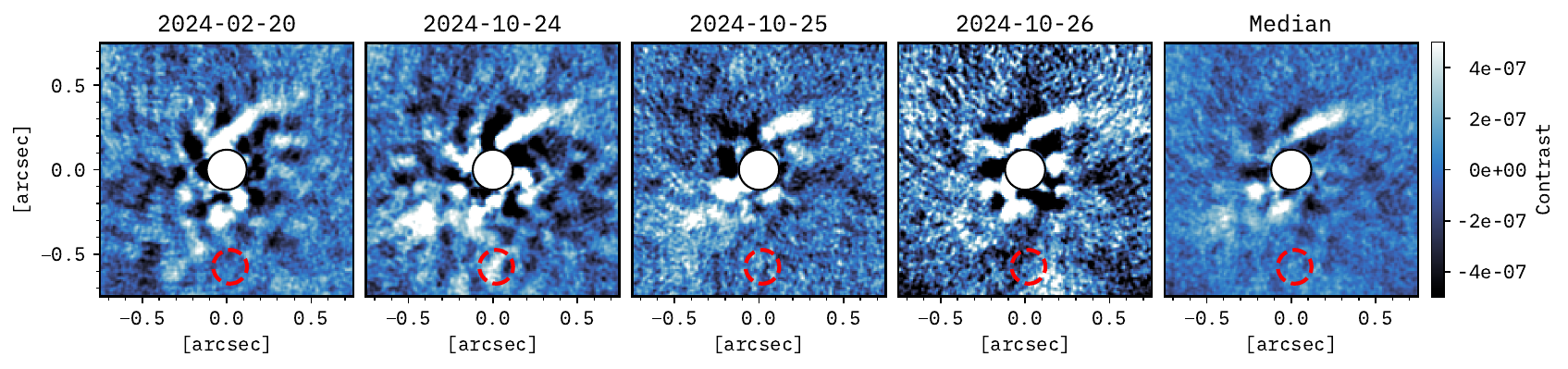}
    \caption{The final AB Aur images and the combined median frame, plotted in units of contrast with north-up and east-right. The images were convolved with a Gaussian kernel of radius 0.75 px to smooth out high spatial frequencies in the frame. The red dashed circle denotes the expected location of the protoplanet candidate.}
    \label{fig:final_midir_reduction_img_multi_plot}
\end{figure*}

\begin{figure}
    \centering
    \includegraphics[width=\linewidth]{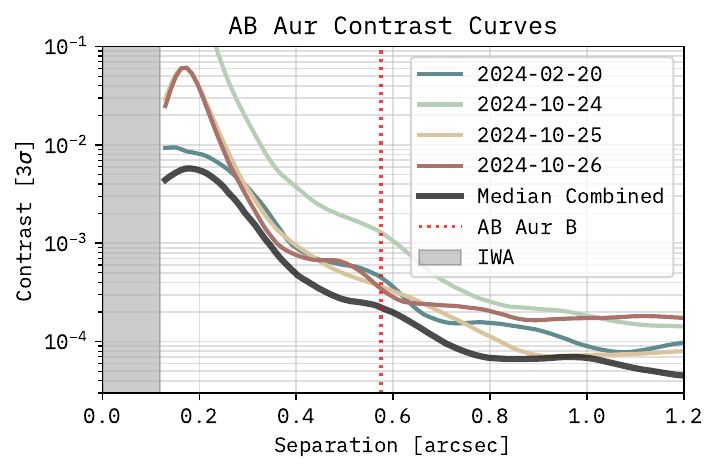}
    \caption{The smoothed contrast curves of our data. The contrast curve uses robust statistics to mitigate the influence of the northern spiral arm and other extended disk features as seen in Figure \ref{fig:final_midir_reduction_img_multi_plot}. Individual contrast curves were calculated by injecting a planet in each data set, then recovering it after the PSF subtraction. The median combined contrast curve was calculated by injecting the same planet in each epoch and recovering it in the final median-combined frame. The expected location of the protoplanet candidate is marked by the dotted line and the inner working angle is marked in grey.}
    \label{fig:final_midir_reduction_ab_aur_cc}
\end{figure}

Our fully processed images are shown in Figure \ref{fig:final_midir_reduction_img_multi_plot} and our resulting contrast curves are shown in Figure \ref{fig:final_midir_reduction_ab_aur_cc}. We reach a contrast of $2\times10^{-4}$ at the expected protoplanet separation, which is likely insufficient  to see the candidate. Models from \cite{choksiSpectralEnergyDistributions2025a} suggest that the contrast is somewhere between $1\times10^{-4}$ and $1\times10^{-5}$ so we therefore take our measurement of $2\times10^{-4}$ to be an upper limit on the brightness of the protoplanet candidate. We do, however, recover the northern spiral arm of the protoplanetary disk around AB Aur and we also observe extended emission to the south-west; both are consistent with intensity maps from \cite{currieImagesEmbeddedJovian2022}. We do not recover much of the spatial structure of the disk at larger separations as these components do not emit strongly in the M-band. Despite varied observing conditions, we obtain similar contrast curves for all of our observations. The disk in each epoch appears visually similar, but the data from 2024-10-26 appears to be the poorest quality, likely due to limited observing time. We note that the sensitivity of the measurement is difficult to judge in images since one needs to perform throughput recovery to get the true contrast curves for the images.

\begin{figure*}
    \centering
    \includegraphics[width=\linewidth]{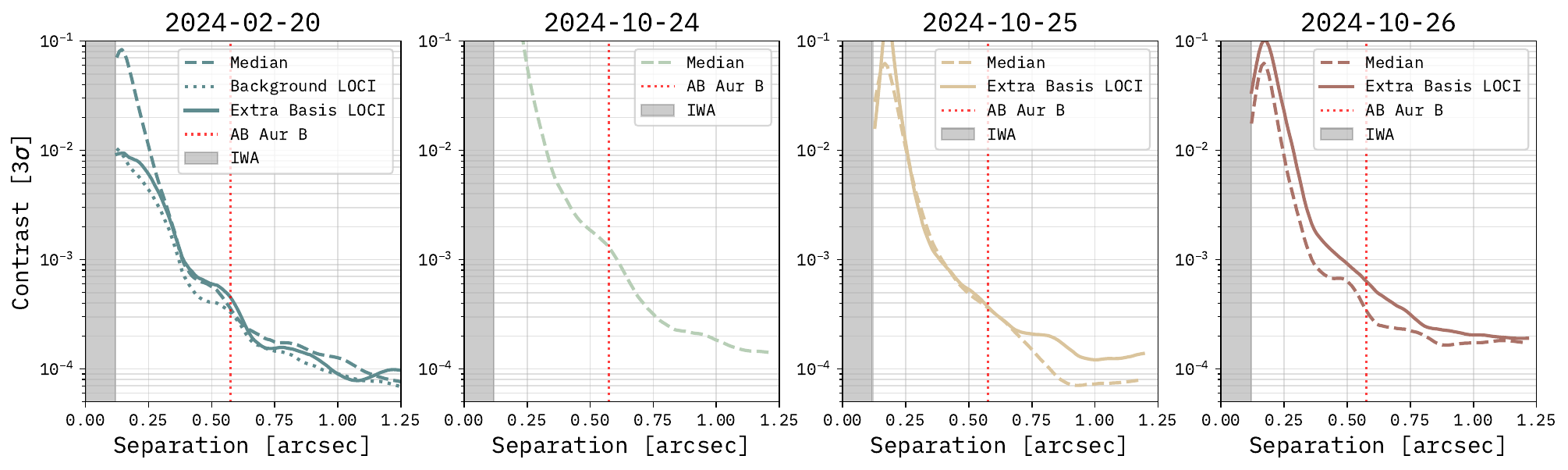}
    \caption{Comparisons of the various techniques for performing background subtraction including simple flat median subtraction, background LOCI, and extra basis LOCI. Not all techniques were possible for each dataset, and only techniques that were possible were applied. The expected location of the protoplanet candidate is marked by the dotted line and the inner working angle is marked in grey.}
    \label{fig:final_midir_reduction_ab_aur_cc_split}
\end{figure*}

Our multi-epoch contrast curve combines the best background subtraction method for each of the epochs, which are background LOCI for the first epoch and median subtraction for the rest. Looking at the contrast curves that compare the different background subtraction techniques in Figure \ref{fig:final_midir_reduction_ab_aur_cc_split} we find that our new technique, Extra Basis LOCI, performed the same or slightly worse than a simple median subtraction. On the first epoch, background LOCI was the best at reducing our data. In this case, background LOCI actually performed better at smaller separation. We find that Extra Basis LOCI likely did not work well with our data since when developing this technique, the nutation behavior was not understood. The idea behind Extra Basis LOCI was that the thermal background was simply a rotated version of itself later during the observation, but Figure \ref{fig:states} shows that that isn't true. At certain elevations when the thermal background changes due to the nutation, higher elevation frames are no longer in the same thermal background state as lower elevation frames and cannot be used for subtraction. In general, significant amounts of K-mirror rotation are required for the PSF to move far away in the K-mirror reference frame so that the background can be measured without its influence. The thermal background noise variation reaches its minimum and transitions to the new state relatively quickly, but these intermediary frames are also show different structure from low and high elevation frames.

The stationary K-mirror epoch, 2024-10-24 also yielded the worst sensitivity out of all the epochs. In turning off the K-mirror, we allowed the PSF to rotate in the frame so that the thermal background would be static. The goal was then to remove the thermal background with just sky frames that we obtain throughout the night. A side effect of keeping the K-mirror static was that the PSF was very unstable throughout the night. The PSF changed significantly as a function of elevation because the AO system uses a different reconstruction matrix depending on the elevation. We therefore did much poorer in our PSF subtraction and get significantly more speckles present in our data. The thermal background was also still difficult to remove since although it no longer rotated, it had variable structure because the noise inversion effect is still visible. Figure \ref{fig:states}, shows how the thermal background states changed that night. This nutation effect limited our thermal background subtraction and the unstable PSF limited our PSF subtraction, leading to significantly worse overall performance during this epoch.

Background LOCI seems to work well in removing the thermal background because it provides continuous information on the background as you dither around the frame. This approach still seems to be the best way to remove the thermal background; however the dithers can be optimized depending on the rate of change of the elevation of the target and how close you are getting to the point of the noise inversion.

Even if we did use a dithering scheme and Background LOCI for all of our observations, it is unlikely we would have detected the candidate protoplanet because all of our epochs (except for the stationary de-rotator) yielded roughly the same contrast at the expected protoplanet separation. To reach deeper contrasts, a coronagraph would need to be used.  However the vortex coronagraph on Keck does not perform well in M-band, and typically suffers from degraded performance due to the thermal background effects as found in \cite{bowens-rubinWolf359Sheeps2023}.

\section{Conclusion}

In this study, we have conducted a thorough evaluation of the thermal background on the Keck II telescope and the systematic error that it introduces in direct imaging observations. From our background-subtracted dataset, we set an upper limit of $2\times10^{-4}$ on the M-band photometry of AB Aur b. The thermal background is surprisingly complex and we believe that we have mostly disentangled the mystery of its exact origin. An excess of blackbody thermal emission from the secondary spiders caused by the nutation in the AO bench optics, coupled with emission from the K-mirror, create a spatially distributed rotating thermal background that is modulated by the deformable mirror in the adaptive optics system. This background is pervasive in M-band observations and its complexity makes it difficult to remove in a model-dependent sense, as we have tried with Extra Basis LOCI. We also find that although the atmosphere is typically invoked as a significant contributor to the thermal background, the large-scale atmospheric state (aside from cloud cover) is not responsible for the systematic component of noise in M-band observations. Additionally, the atmospheric properties within the dome and close to the instruments are not the main drivers of the thermal background. Despite our best attempts at developing new techniques for removing this background, Background LOCI remains the best approach for combating the thermal background.

The applications of this study are limited to NIRC2 M-band observations on Keck, but other studies \citep{sauterDetectionLimitsThermalinfrared2024a, hunzikerPCAbasedApproachSubtracting2018, burtscherPhysicalUnderstandingThermal2020} have shown that there is a complex thermal background present on many other telescopes and instruments. More often than not, the thermal background is largely influenced by the optics of the telescope themselves rather than atmospheric variation. 

In the near future, 30-m class telescopes will bring an expanded mid-IR imaging capabilities (\SI{5}{}-\SI{25}{\micro\meter}) to the ground with instruments such as MICHI  \citep{packhamKeyScienceDrivers2012} on the TMT and MICADO \citep{daviesMICADOFirstLight2016} on the ELT. These instruments will take advantage of the relatively high angular resolution of ground-based telescopes at these wavelengths. The thermal background environment on these telescopes and instruments will present significant engineering challenges that will be difficult to predict before going on sky. A deep dive into the systematics of the thermal background on those telescopes, like this study, will inevitably need to be conducted to achieve the science goals of these instruments.

\section{Acknowldegements}
The authors would like to thank Jim Lyke for his helpful discussion on determining the origin of the noise components of the background. This research has made use of the Keck Observatory Archive (KOA), which is operated by the W. M. Keck Observatory and the NASA Exoplanet Science Institute (NExScI), under contract with the National Aeronautics and Space Administration. The authors wish to recognize and acknowledge the very significant cultural role and reverence that the summit of Maunakea has always had within the Native Hawaiian community. We are most fortunate to have the opportunity to conduct observations from this mountain. 

\software{ADI.jl \citep{Lucas2020}, Astropy \citep{astropy:2013, astropy:2018, astropy:2022}, NumPy \citep{harris2020array}, SciPy \citep{2020SciPy-NMeth}, Matplotlib \citep{Hunter:2007}}

\bibliography{main}{}
\bibliographystyle{aasjournalv7}

\end{document}